# Large-Area Synthesis of High-Quality and Uniform Graphene Films on Copper Foils

Xuesong Li[a], Weiwei Cai[a], Jinho An[a], Seyoung Kim[b], Junghyo Nah[b], Dongxing Yang[a], Richard Piner[a], Aruna Velamakanni[a], Inhwa Jung[a], Emanuel Tutuc[b], Sanjay K. Banerjee[b], Luigi Colombo[c*], Rodney S. Ruoff[a*]

[a] Department of Mechanical Engineering and the Texas Materials Institute, 1 University Station C2200, The University of Texas at Austin, Austin, TX 78712-0292
[b] Department of Electrical and Computer Engineering, Microelectronics Research Center, The University of Texas at Austin, Austin, Texas 78758, USA
[c] Texas Instruments Incorporated, Dallas, TX 75243
*To whom correspondence should be addressed: r.ruoff@mail.utexas.edu, colombo@ti.com

Abstract

Graphene has been attracting great interest because of its distinctive band structure and physical properties. Today, graphene is limited to small sizes because it is produced mostly by exfoliating graphite. We grew large-area graphene films of the order of centimeters on copper substrates by chemical vapor deposition using methane. The films are predominantly single layer graphene with a small percentage (less than 5%) of the area having few layers, and are continuous across copper surface steps and grain boundaries. The low solubility of carbon in copper appears to help make this growth process self-limiting. We also developed graphene film transfer processes to arbitrary substrates, and dual-gated field-effect transistors fabricated on Si/SiO$_2$ substrates showed electron mobilities as high as 4050 cm$^2$V$^{-1}$s$^{-1}$ at room temperature.

Graphene, a monolayer of $sp^2$-bonded carbon atoms, is a quasi-two-dimensional (2D) material. Graphene has been attracting great interest because of its distinctive band structure and physical properties (*1*). Today, the size of graphene films produced is limited to small sizes (usually < 1000 μm$^2$) because the films are produced mostly by exfoliating graphite, which is not a scalable technique. Graphene has also been synthesized by the desorption of Si from SiC single crystal surfaces which yields a multilayered graphene structure that behaves like graphene (*2, 3*), and by a surface precipitation process of carbon in some transition metals (*4-8*).

Electronic application will require high-quality large area graphene that can be manipulated to make complex devices and integrated in silicon device flows. Field effect transistors (FETs) fabricated with exfoliated graphite have shown promising electrical properties (*9, 10*), but these devices will not meet the silicon device scaling requirements, especially those for power reduction and performance. One proposed device that could meet the silicon roadmap requirements beyond the 15 nm node by Banerjee *et al.* (*11*) The device is a 'BisFET' (bilayer pseudospin FET) device which is made up of two graphene layers separated by a thin dielectric. The ability to create this device can be facilitated by the availability of large-area graphene. Making a transparent electrode, another promising application of graphene, also requires large films (*6, 12-14*).

At this time, there is no pathway for the formation of a graphene layer that can be exfoliated from or transferred from the graphene synthesized on SiC, but there is a way to grow and transfer graphene grown on metal substrates (*5-7*). Although graphene has been grown on a number of metals, we still have the challenge of growing large-area graphene. For example, graphene grown on Ni seems to be limited by its small grain size, presence of multilayers at the grain boundaries, and the high solubility of carbon (*6, 7*). We have developed a graphene chemical vapor deposition (CVD) growth process on copper foils (25 μm thick in our experiment). The films grow directly on the surface by a surface catalyzed process and the film is predominantly graphene with <5% of the area having two- and three-layer graphene flakes. Under our processing conditions, the two- and three-layer flakes do not grow larger with time. One of the major benefits of our process is that it can be used to grow graphene on 300 mm copper films on Si substrates (a standard process in Si technology). It is also well known that annealing of Cu can lead to very large grains.

As described in (*15*), we grew graphene on copper foils at temperatures up to 1000 °C by CVD of carbon using a mixture of methane and hydrogen. Figure 1A shows a scanning electron microscopy (SEM) image of graphene on a copper substrate where the Cu grains are clearly visible. A higher-resolution image of graphene on Cu (Fig. 1B) shows the presence of Cu surface steps, graphene "wrinkles", and the presence of non-uniform dark flakes. The wrinkles associated with the thermal expansion coefficient difference between Cu and graphene are also found to cross Cu grain boundaries, indicating that the graphene film is continuous. The inset in Fig.1b shows transmission electron microscopy (TEM) images of graphene and bilayer graphene. With the use of a process similar to that described in ref. (*16*), the as-grown graphene can be easily transferred to alternative substrates such as $SiO_2$/Si or glass (Figs. 1, C and D), for further evaluation and for various applications; a detailed transfer process is described in the supplemental section. The process and method used to transfer graphene from Cu was the same for the $SiO_2$/Si substrate and the glass substrate. Although it is difficult to see the graphene on the $SiO_2$/Si substrate, a similar graphene film from another Cu substrate transferred on glass clearly shows that it is optically uniform.

We used Raman spectroscopy to evaluate the quality and uniformity of graphene on $SiO_2$/Si substrate. Figure 2 shows SEM and optical images with the corresponding Raman spectra and maps of the D, G and 2D bands providing information on the defect density and film thickness. The Raman spectra are from the spots marked with the corresponding colored circles shown in the other panels (in Figs. 2, A and B, green arrows are used instead of circles so as to show the trilayer region more clearly). The thickness and uniformity of the graphene films were evaluated via color contrast under optical microscope (*17*) and Raman spectra (*7, 18, 19*). The Raman spectrum from the lightest pink background in Fig. 2B shows typical features of monolayer graphene, e.g., ~0.5 G-to-2D intensity ratio, and a symmetric 2D band centered at ~2680 $cm^{-1}$ with a full-width of half-maximum (FWHM) of ~33 $cm^{-1}$. The second lightest pink "flakes" (blue circle) correspond to bilayer graphene and the darkest one (green arrow) represents trilayer graphene. This thickness variation is more clearly shown in the SEM image in Fig. 2A. The D map in Fig. 2D, which has been associated with defects in graphene, is rather uniform and near the background level, except for regions where wrinkles are present and close to few-layer regions. The G and the 2D maps clearly show the presence of more than one layer in the flakes. In the wrinkled regions, there are peak height variations in both the G and 2D bands, and there is a broadening of the 2D band. An analysis of the intensity of the optical image over the whole sample (1 cm by 1 cm) showed that the area with the lightest pink color is more than 95%, and all 40 Raman spectra randomly collected from this area show monolayer graphene. There is only a small fraction of trilayer or few-layer (<10) graphene (<1%) and the rest is bilayer graphene (~ 3-4%).

We grew films on Cu as a function of time and Cu foil thickness under isothermal and isobaric conditions. Using the process flow described in (*15*) we found that graphene growth on Cu is self-limited; growth that proceeded for more than 60 min yielded a similar structure to growth runs performed for ~10 min. For times much less than 10 min, the Cu surface is usually not fully covered [SEM images of graphene on Cu with different growth time are shown in figure S3 (*15*)]. The growth of graphene on Cu foils of varying thickness (12.5, 25, and 50 μm) also yielded similar graphene structure with regions of double and triple flakes but neither discontinuous monolayer graphene for thinner Cu foils nor continuous multilayer graphene for thicker Cu foils, as we would have expected based on the precipitation mechanism. According to these observations, we concluded that graphene is growing by a surface-catalyzed process rather than a precipitation process as reported by others for Ni (*5-7*). Monolayer graphene formation caused by surface segregation or surface adsorption of carbon has also been observed on transition metals such as Ni and Co at elevated temperatures by Blakely and coauthors (*20-22*). However, when the metal substrates were cooled down to room temperature, thick graphite films were obtained because of precipitation of excess C from these metals, in which the solubility of C is relatively high.

In recent work, thin Ni films and a fast-cooling process have been used to suppress the amount of precipitated C. However, this process still yields films with a wide range of graphene layer thicknesses, from one to a few tens of layers and with defects associated with fast cooling (*5-7*). Our results suggest that the graphene growth process is not one of C precipitation but rather a CVD process. The precise mechanism will require additional experiments to understand in full, but very low C solubility in Cu (*23-25*), and poor C saturation as a result of graphene surface coverage may be playing a role in limiting or preventing the precipitation process altogether at high temperature, similar to the case of impeding of carburization of Ni (*26*). This provides a pathway for growing self-limited graphene films.

To evaluate the electrical quality of the synthesized graphene, we fabricated dual-gated FET with $Al_2O_3$ as the gate dielectric and measured them at room temperature. Along with a device model that incorporates a finite density at the Dirac point, the dielectric, and the quantum capacitances (*9*), the data are shown in Fig. 3. The extracted carrier mobility for this device is ~4050 cm$^2$V$^{-1}$s$^{-1}$, with the residual carrier concentration at the Dirac point of $n_0 = 3.2 \times 10^{11}$ cm$^{-2}$. These data suggest that the films are of reasonable quality, at least sufficient to continue improving the growth process to achieve a material quality equivalent to the exfoliated natural graphite.

**Supporting Information Available:** Materials and Methods. Fig. S1, S2, and S3.

Figure Captions

Fig.1. (A) SEM image of graphene on a copper foil with a growth time of 30 min. (B) High-resolution SEM image showing a Cu grain boundary and steps, two- and three-layer graphene flakes, and graphene wrinkles. Inset in (B) shows TEM images of folded graphene edges. (C and D) Graphene films transferred onto a $SiO_2$/Si substrate and a glass plate, respectively.

Fig.2. (A) SEM image of graphene transferred on $SiO_2$/Si (285-nm thick oxide layer) showing wrinkles, and 2 and 3 layer regions. (B) Optical microscope image of the same regions as (A). (C) Raman spectra from the marked spots with corresponding colored circles or arrows showing the presence of graphene, 2 layers of graphene and 3 layers of graphene; (D, E, and F) Raman maps of the D (1300 to 1400 $cm^{-1}$), G (1560 to 1620 $cm^{-1}$), and 2D (2660 to 2700 $cm^{-1}$) bands, respectively (WITec alpha300, $\lambda_{laser}$ = 532 nm, ~500 nm spot size, 100× objector). Scale bars are 5 μm.

Fig.3. (A) Optical microscope image of a graphene FET. (B) Device resistance vs top-gate voltage ($V_{TG}$) with different back-gate ($V_{BG}$) biases and vs $V_{TG}$-$V_{\_Dirac,TG}$ ($V_{TG}$ at the Dirac point), with a model fit (solid line).

Figures

Fig. 1

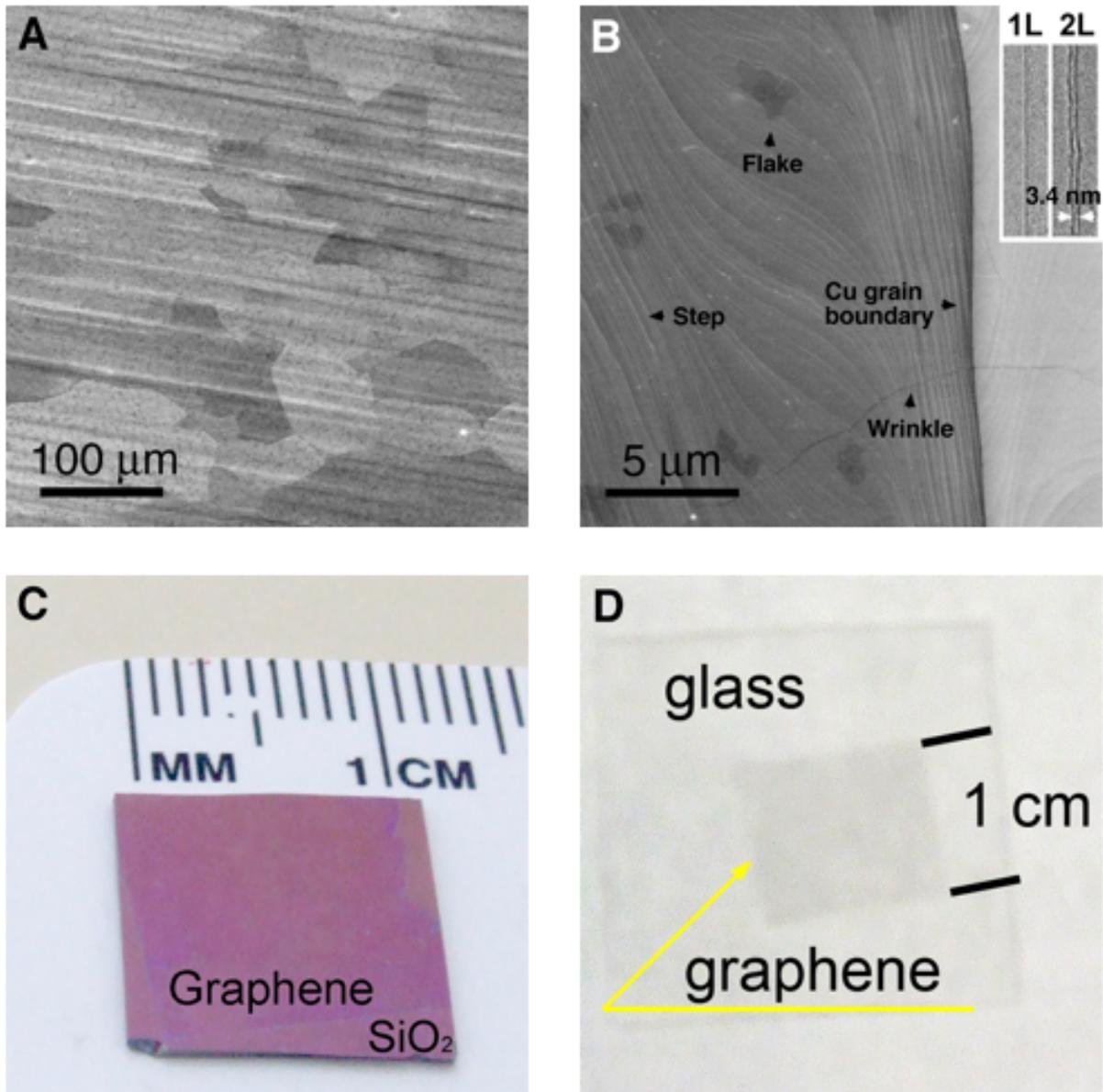

Fig. 2

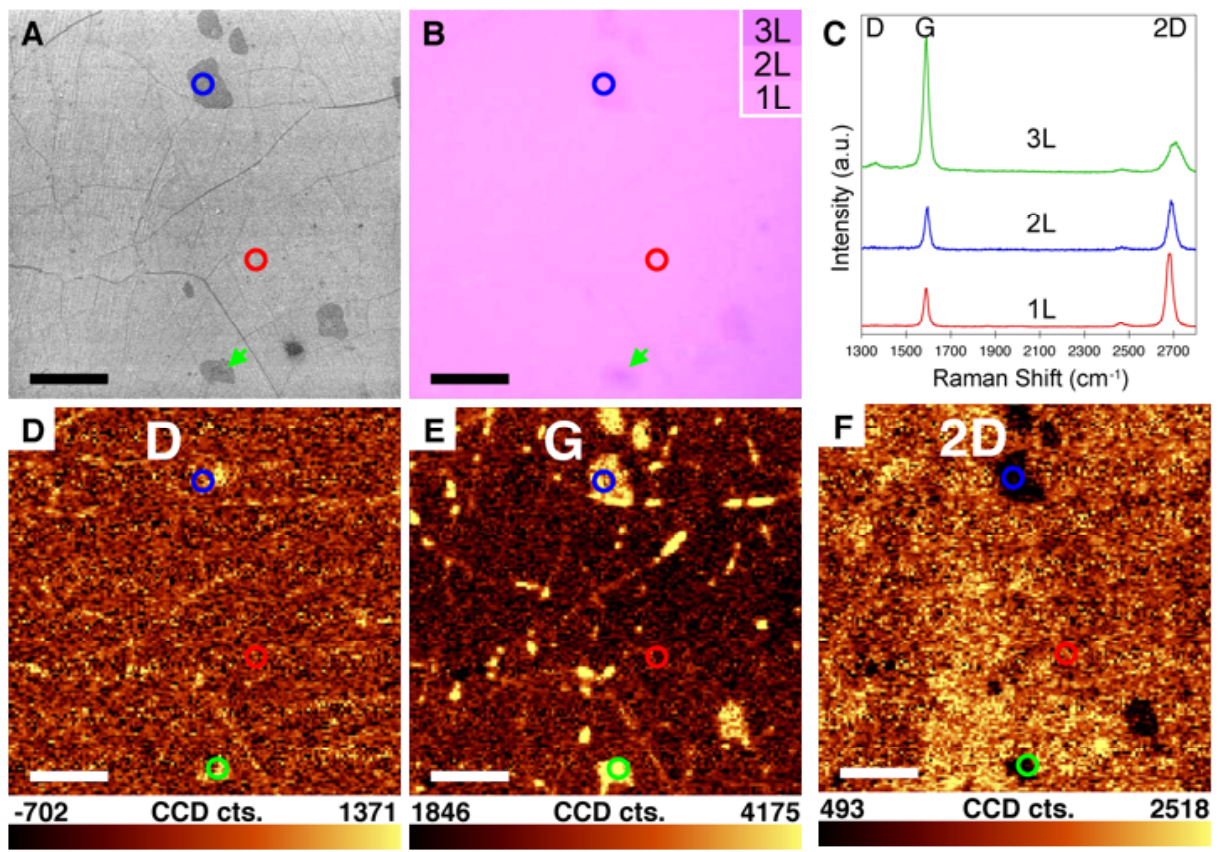

Fig. 3

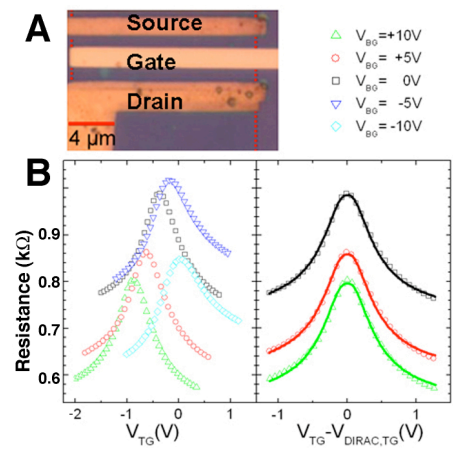

Supporting Online Material for:

# Large Area Synthesis of High-Quality and Uniform Graphene Films on Copper Foils

Xuesong Li[a], Weiwei Cai[a], Jinho An[a], Seyoung Kim[b], Junghyo Nah[b], Dongxing Yang[a], Richard Piner[a], Aruna Velamakanni[a], Inhwa Jung[a], Emanuel Tutuc[b], Sanjay K. Banerjee[b], Luigi Colombo[c*], Rodney S. Ruoff[a*]

[a] Department of Mechanical Engineering and the Texas Materials Institute, 1 University Station C2200, The University of Texas at Austin, Austin, TX 78712-0292
[b] Department of Electrical and Computer Engineering, Microelectronics Research Center, The University of Texas at Austin, Austin, Texas 78758, USA
[c] Texas Instruments Incorporated, Dallas, TX 75243
*Correspondence to: r.ruoff@mail.utexas.edu, colombo@ti.com

**Materials and Methods**

*Growth and transfer of graphene films*

Graphene films were primarily grown on 25-μm thick Cu foils (Alfa Aesar, item No. 13382, cut into 1 cm strips) in a hot wall furnace consisting of a 22-mm ID fused silica tube heated in a split tube furnace; several runs were also done with 12.5- and 50-μm thick Cu foils (also from Aesar). A typical growth process flow is: (1) load the fused silica tube with the Cu foil, evacuate, back fill with hydrogen, heat to 1000 °C and maintain a $H_2(g)$ pressure of 40 mTorr under a 2 sccm flow; (2) stabilize the Cu film at the desired temperatures, up to 1000 °C, and introduce 35 sccm of $CH_4(g)$ for a desired period of time at a total pressure of 500 mTorr; (3) after exposure to $CH_4$, the furnace was cooled to room temperature. The experimental parameters (temperature profile, gas composition/flow rates, and system pressure) are shown in Fig. S1. The cooling rate was varied from > 300 °C/min to about 40 °C/min which resulted in films with no discernable differences. Fig. S2 shows the Cu foil with the graphene film, compared to the as-received Cu foil.

Graphene films were removed from the Cu foils by etching in an aqueous solution of iron nitrate. The etching time was found to be a function of the etchant concentration, the area, and thickness of the Cu foils. Typically, a 1 cm$^2$ by 25-μm thick Cu foil can be dissolved by a 0.05 g/ml iron nitrate solution over night. Since graphene grows on both sides of the Cu foil, two films are exfoliated during the etching process. We used two methods to transfer the graphene from the Cu foils: (1) after the copper film is dissolved, a substrate is brought into contact with the graphene film and it is 'pulled' from the solution; (2) the surface of the graphene-on-Cu is coated with polydimethylsiloxane (PDMS) or poly-methyl methacrylate (PMMA) and after the Cu is dissolved the PDMS-graphene is lifted from the solution, similar to the method reported in the reference metioned in the main text. The first method is simple, but the graphene films break and tear more readily. The graphene films are easily transferred with the second method to other desired substrates such as $SiO_2/Si$, with significantly fewer holes or cracks (< 5% of the film area).

**Figures**

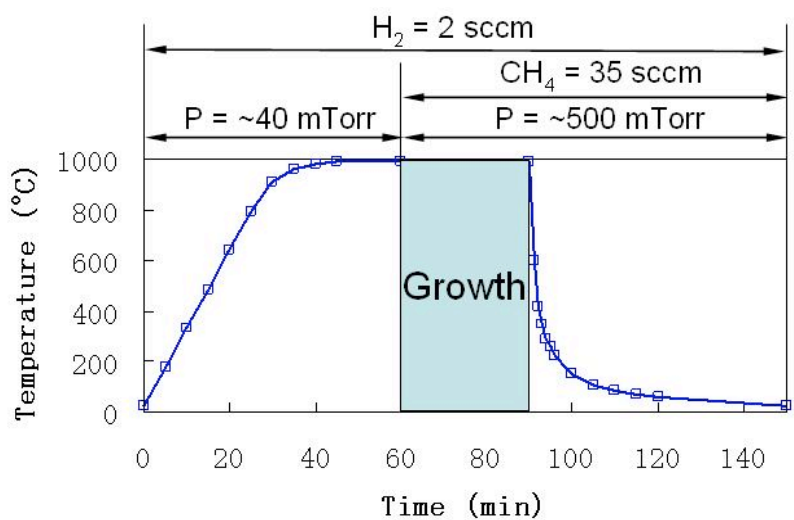

Figure S1. Time dependence of experimental parameters: temperature, pressure, and gas composition/flow rate.

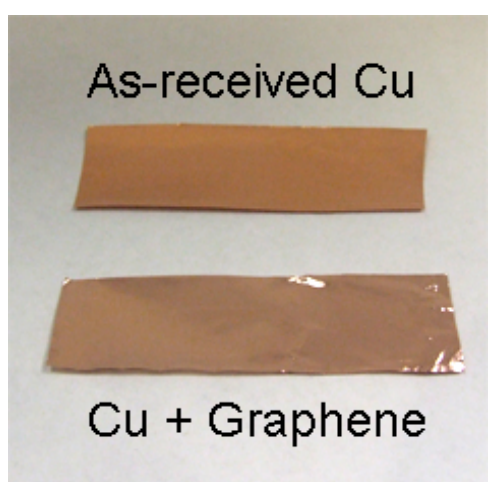

Figure S2. Photos of as-received Cu foil, and Cu foil covered with graphene. The Cu foil with graphene has a smooth surface and is "shinier" compared to the as-received Cu foil, which has a thin but rough oxide layer.

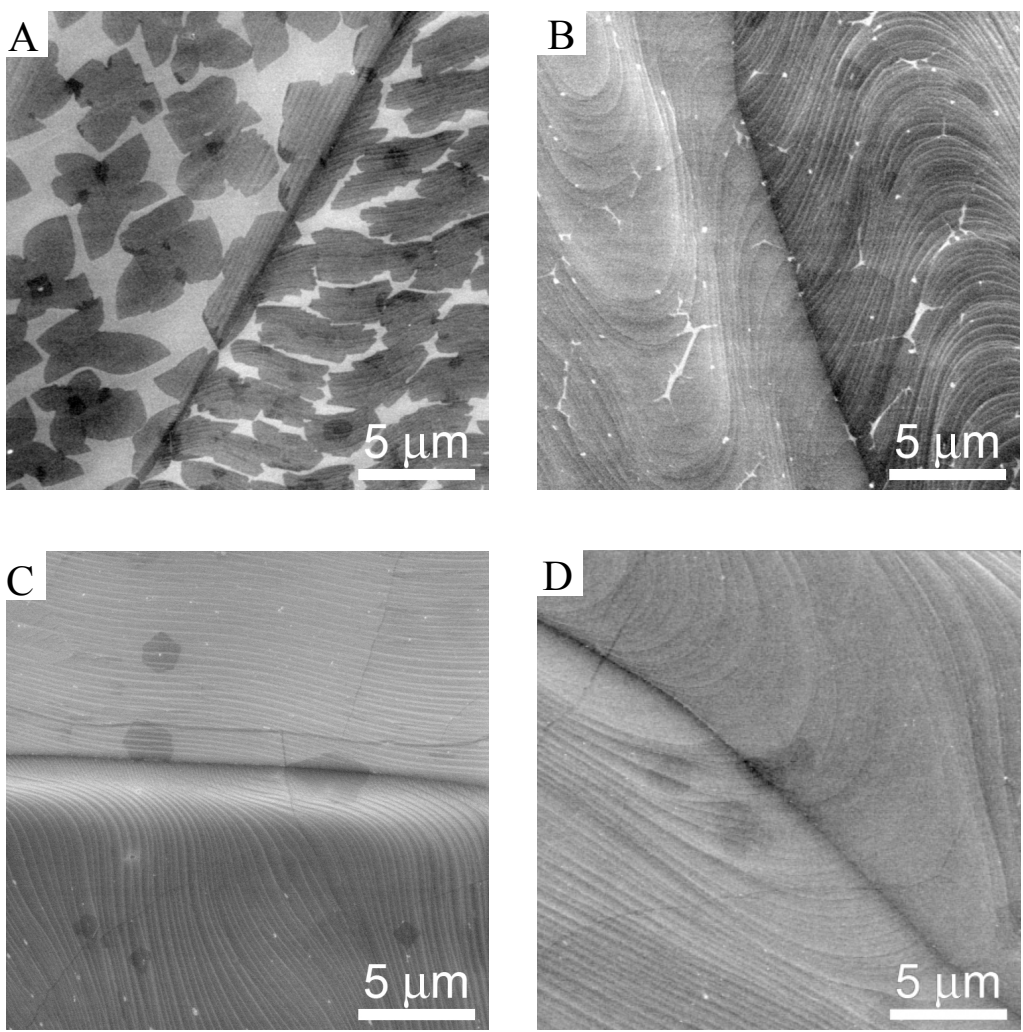

Figure S3. SEM images of graphene on Cu with different growth times of (A) 1 min, (B) 2.5 min, (C) 10 min, and (D) 60 min, respectively.

**Supporting Online Material**
www.sciencemag.org
Materials and Methods
Figs. S1, S2, S3